# Doping-tunable Fermi surface with persistent topological Hall effect in axion candidate EuIn$_2$As$_2$

Jian Yan,[1,*] Jianguo Si,[2] Zhongzhu Jiang,[3,4] Hanming Ma,[1] Yoshiya Uwatoko,[1] Bao-Tian Wang,[5] Xuan Luo,[3] Yuping Sun,[6,3,7,†] and Minoru Yamashita[1,§]

[1] *The Institute for Solid State Physics, The University of Tokyo, Kashiwanoha 5-1-5, Kashiwa, Chiba 277-8581, Japan*

[2] *Songshan Lake Materials Laboratory, Dongguan, Guangdong 523808, China*

[3] *Key Laboratory of Materials Physics, Institute of Solid State Physics, HFIPS, Chinese Academy of Sciences, Hefei, 230031, China*

[4] *University of Science and Technology of China, Hefei, 230026, China*

[5] *Institute of High Energy Physics, Chinese Academy of Sciences, Beijing 100049, China*

[6] *Anhui Province Key Laboratory of Low-Energy Quantum Materials and Devices, HFIPS, High Magnetic Field Laboratory, Chinese Academy of Sciences, Hefei, 230031, China*

[7] *Collaborative Innovation Center of Advanced Microstructures, Nanjing University, Nanjing, 210093, China*

[*]jianyan@issp.u-tokyo.ac.jp
[†]ypsun@issp.ac.cn
[§]my@issp.u-tokyo.ac.jp

abstract
**Abstract**
Rare-earth Zintl compound EuIn$_2$As$_2$ has been theoretically recognized as a candidate for realizing an intrinsic antiferromagnetic (AFM) bulk axion insulator and a higher-order topological state, which provides a fertile platform to explore novel topological transport phenomena. However, the axion state has yet to be realized because EuIn$_2$As$_2$ is highly hole-doped. Here, we synthesized a series of high-quality Ca-doped EuIn$_2$As$_2$ (Ca$_x$Eu$_{1-x}$In$_2$As$_2$, $x = 0 \sim 0.25$) single crystals to tune the Fermi energy above the hole pocket. Our Hall measurements reveal that the isovalent Ca substitution decreases the hole carrier density by shrinking the lattice spacing, which is also confirmed by our first-principles calculations. We further find that both the temperature dependence of the magnetic susceptibility with a local maximum at the Néel temperature and the topological Hall effect originating from the finite real-space spin chirality persist in the Ca-doped samples as observed in the pristine EuIn$_2$As$_2$, despite that the nonmagnetic Ca substitution decreases the effective moment and the Néel temperature. These results show that the Ca substitution tunes the Fermi energy while keeping the AFM magnetic structure, suggesting that the axion insulating state may be realized by further Ca substitution.

**Introduction**

Magnetic topological materials, which represent a class of compounds with properties that are strongly influenced by the topology of their electronic wavefunctions coupled with the magnetic spin configuration, have attracted much attention due to their usefulness in future applications [1-5]. Such materials provide a unique playground for exploring exotic quantum phenomena in condensed matter physics, such as the axion insulator state and the quantum anomalous Hall effect (QAHE). Currently, intrinsically antiferromagnetic (AFM) topological insulators (TIs) are intensely studied to explore such exotic quantum phenomena [5]. In recent years, materials from the homologous series of the MnBi$_{2n}$Te$_{3n+1}$ family have been intensively studied as intrinsically magnetic and stoichiometric TIs [4, 6, 7]. An intrinsic van der Waals AFM with a nontrivial surface state was first identified in MnBi$_2$Te$_4$ by first-principles density functional theory (DFT) calculations and experiments [6]. Recently, Deng et al. have reported QAHE in 5-setuplayers (SLs) thin film with uncompensated AFM moments at 1.4 K [7] and axion insulator state in 6-SLs thin film with fully compensated antiferromagnetism [4]. However, the study of AFM TIs is still in the early stage, studies in other candidate materials are desired.

Beyond the MnBi$_2$Te$_4$ family, rare-earth Zintl compound EuIn$_2$As$_2$, another unique natural heterostructure consisting of both topological InAs layers and magnetic Eu interlayer, has been suggested as another promising intrinsic magnetic TI by first-principles DFT calculations [8]. It has been predicted that the in-plane A-type AFM order in EuIn$_2$As$_2$ hosts a bulk axion insulator phase. Further, by orienting the magnetic moment of the A-type AFM order to the *c*-axis, a higher-order topological state can be realized [8]. Such an easy-tunable topological state by the magnetism in EuIn$_2$As$_2$ offers a unique opportunity to explore novel topological states. However, previous transport and angle-resolved photoemission spectroscopy measurements have shown that EuIn$_2$As$_2$ is highly hole-doped [9-11], which obstructs to realize the axion insulator nature. Chemical doping has been attempted to compensate the hole carrier, which is however often accompanied by unintended changes in the magnetic property. For example, P-doping in EuIn$_2$As$_2$ changes the AFM order into a spin glass state, as well as the disappearance of the Néel peak in the temperature dependence of the magnetic susceptibility [12]. Also, Ba substitution in the magnetic Weyl semimetal EuCd$_2$As$_2$ tilts the in-plane magnetic moments of the AFM order [13].

In this paper, we conducted detailed transport measurements on a series of Ca-doped EuIn$_2$As$_2$. Firstly, the *a*- and *c*-axis lengths decrease with increasing doping ratios, indicating a lattice contraction. We find that this lattice contraction by the isovalent Ca doping effectively decreases the hole carrier, as confirmed by our first-principles calculations. Although the magnetic transition temperature decreases with increasing *x*, both the temperature dependence of the magnetic susceptibility and the topological Hall effect are observed in all the Ca-doped samples as observed in the pristine EuIn$_2$As$_2$ [11], suggesting that the AFM magnetic structure remains almost the same. Our results indicate that nonmagnetic Ca element doping can modify the band structure without changing the magnetic state of EuIn$_2$As$_2$, providing an opportunity to realize the axion insulating state by only tuning the carrier density.

**Experimental methods**

Ca-doped EuIn$_2$As$_2$ single crystals were grown by the self-flux method similar to previous reports [11, 12]. A molar ratio of Ca:Eu:In:As = *x*:1-*x*:12:3 was used, where the large excess In serves as the flux. All reagents were ground into alumina crucible and sealed in a quartz ampoule under high vacuum and all were done in an Ar-filled glove box. The sealed quartz ampoule was heated to 1050°C and maintained for 12 hours, then cooled down to 700-600°C over 100 hours for different doping ratios. At this temperature, the quartz ampoule was quickly taken out from the furnace and decanted with a centrifuge to separate Ca-doped EuIn$_2$As$_2$ single crystals.

The elemental doping composition was measured by energy dispersive *x*-ray spectroscopy (EDS, Oxford X-Max 50). Powder *x*-ray diffraction (XRD) at 293(2) K was performed using a Rigaku XtaLab MicroMax007 HFMR *x*-ray source [Mo K α radiation (λ = 0.71073Å)] with a HyPix-6000 detector. Using OLEX2 [14], the structure was solved by direct methods (SHELXT2018/2) [15] and refined by full-matrix least-squares methods on F$^2$ values (SHELXL 2018/3) [15]. Magnetization and electrical transport measurements were carried out by using a Quantum Design Magnetic Property Measurement System (MPMS-XL5) and a Physical Properties Measurement System for (PPMS-9 T) 2 K < *T* < 400 K and $\mu_0 H$ < 9 T. Standard four-probe and five-probe methods were used for the longitudinal resistivity and the Hall measurements with a current in the *ab* plane, respectively. All the Hall resistivity were obtained by $\rho_{xy}(\mu_0 H) = [\rho_{xy}(+\mu_0 H) - \rho_{xy}(-\mu_0 H)]/2$ to remove the influence of misalignment of the contacts.

**Calculation details**

The band structure calculations were performed by the Vienna Ab-initio Simulation Package (VASP), which is based on DFT [16, 17], where the spin-orbit coupling effect was considered. According to the generalized gradient approximation, the projector-augmented wave pseudopotential was used to treat the interaction between the valance electrons and ionic cores [18], which is parametrized by Perdew-Burke-Ernzerhof functions [19]. The on-site Coulombic interaction was introduced by the DFT+$U$ method. According to the previous reports [8, 10, 20], the effective Hubbard $U$ was set as 5 eV for the strongly localized Eu 4$f$ orbitals. The cut-off energy for the wave function is set as 300 eV. To evaluate the effect of Ca doping, a 2 × 2 supercell was created. The Brillouin zone was sampled by using a G-centered mesh with a sampling density of 9 × 9 × 3. All atoms were fully relaxed until all the residual forces were less than 0.001 eV/Å. The data post processing and band unfolding were obtained by Vaspkit code [21].

**Results and discussion**

EuIn$_2$As$_2$ has a three-dimensional structure and crystallizes in the hexagonal $P6_3/mmc$ (No. 194) space group, with alternating stacking of Eu$^{2+}$ and [In$_2$As$_2$]$^{2-}$ layers along the $c$ axis [9]. Figure 1(a) shows the crystal structure of Ca-doped EuIn$_2$As$_2$, with Ca substituting Eu sites. The elemental compositions of all single crystals were measured by EDS and the results are consistent with the formation of Eu-122 phase. We find that the absorption peak of Ca in the EDS spectrum significantly overlaps with the In peak, causing a large ambiguity in the estimation of the Ca concentration from the EDS results. Therefore, all values of $x$ quoted in this paper refer to the nominal Ca ratio determined by the ratio Ca to Eu of the starting materials. It should be noted that the EDS results also reveal a nearly homogeneous distribution of $x$ for these crystals and no possible phase separations could be identified, as shown in Fig. S1 and Table I in Supplementary Information [22]. Single crystals with five typical doping concentrations ($x$ = 0, 0.06, 0.12, 0.2 and 0.25) were selected for further measurements. Unfortunately, until now no high-quality single crystal has been obtained for a higher Ca ratio ($x > 0.25$) using the current synthesis technique. Figure 1(b) shows the powder XRD patterns of these single crystals. The powder XRD data was collected on crushed Ca$_x$Eu$_{1-x}$In$_2$As$_2$ single crystals at room temperature. We also present the structural Rietveld refinement profiles of the XRD data (Fig. S2). The XRD refinement result shows no structure changes in Ca doping, and no obvious impurity phase is observed, assuring the intrinsic nature of doping dependent physical properties in our experiments. The refined lattice parameters $a$ and $c$ are plotted against Ca ratio $x$ in Figure 1(c). The blue line in Figure 1(c) indicates the extrapolation of both lattice parameters from EuIn$_2$As$_2$ to CaIn$_2$As$_2$. With the increase of $x$, both $a$ and $c$ decrease along the extrapolation line, showing that the substitution of Eu$^{2+}$ by Ca$^{2+}$ leads to a lattice contraction as expected by the EDS results. The decrease of $a$ and $c$ are further evidenced by the systematic shift of (107) peak to a higher angle with increasing $x$, as shown in Figure 1(d).

Next, we present the evolution of the electronic transport properties with Ca doping in $Ca_xEu_{1-x}In_2As_2$. Figure 2(a) shows the temperature dependence of the longitudinal in-plane resistivity $\rho_{xx}(T)$ of $Ca_xEu_{1-x}In_2As_2$ single crystals, which are normalized by the resistivity at 200 K. In the pristine sample [11], $\rho_{xx}(T)$ shows a metallic temperature dependence down to ~50 K, which is followed by a sharp peak at the Néel temperature $T_N$ caused by the decrease of the magnetic scattering. All the Ca-doped samples also show the metallic temperature dependence with suppressed $T_N$ which is determined by the zero-crossing point of the derivative of the resistivity with respect to temperature ($d\rho/dT$) as marked by the arrows in the inset of Fig. 2(a). As listed in Table II, we find that $T_N$ is monotonically lowered as increasing the Ca doping. We also find that the Ca doping enhances the magnetic scattering on the conduction electrons as shown by the increase of the residual resistivity at 2 K (Table II) except the samples of $x$ = 0.06 and 0.2. In these samples, the decrease of $\rho_{xx}$ from the room temperature to $T_N$ is larger and the peak of $\rho_{xx}$ at $T_N$ is smaller than those in other doped samples, implying smaller magnetic scattering effects in the two samples. This different temperature dependence might be caused by a partial ordering of the Ca dopants by the different growth temperatures during the sample synthesis. More advanced experiments, such as other synthesis techniques or synthesizing small size of thin films, are necessary for controlling the homogeneity by tuning the doping sites precisely.

To further study the Ca-doping effect, we investigate the temperature dependence of the magnetic susceptibility ($\chi(T)$) at a magnetic field of 0.01 T applied along the $c$ axis. As shown in Fig. 2(b), the temperature dependence of $\chi(T)$ of all the Ca-doped samples has a local maximum at around $T_N$ determined by $d\rho/dT$. We find that the gradual decrease of $T_N$ determined by the peak in $\chi(T)$ agrees well with that determined by $d\rho/dT$. In addition, no difference in $\chi(T)$ measured in the zero-field cool (ZFC) and field cool (FC) is observed in all the Ca-doped samples, showing the absence of a spin-glass state. These results suggest the persistence of the AFM state with the suppressed $T_N$ in the Ca-doped samples, contrary to the change of the AFM state to the spin glass state observed in the P-doped samples [12]. We note that the non-monotonic dependence of the absolute value of $\chi$ on the Ca-doping ratio $x$ might be caused by effects of magnetic impurities that affect the magnitude of $\chi$ at low temperatures and low magnetic fields.

The temperature dependence of $\chi(T)$ above $T_N$ follows the Cuire-Weiss law, $\chi(T) = \frac{C}{T-T_\theta}$, where $T_\theta$ is the Curie-Weiss temperature and $C$ is the Curie constant given by the effective moments $\mu_{eff}$ (Fig. S3), [23]. In Eu-122 system, the magnitude of the magnetic susceptibility consists only of the Eu moments with the effective moment $g\sqrt{S(S+1)} = 7.94\mu_B$ with $S = 7/2$, which well agrees with $\mu_{eff}$ of the pristine sample (8.02 $\mu_B$). As listed in Table II of the Supplementary Information [22], both $T_\theta$ and $\mu_{eff}$ determined by the Curie-Weiss fit are suppressed by the Ca doping, showing the dilution effect on the Eu moments as expected. The positive $T_\theta$ of all the samples indicates the presence of a dominant ferromagnetic (FM) interaction mixed with an AFM interaction forming the AFM order. It should be noted that the previous study under the physical pressure shows that the lattice contraction in EuIn2As2 results in the increase of $T_\text{N}$ [24], in contrast to the suppression of $T_\text{N}$ in the Ca-doped samples. This enhancement of $T_\text{N}$ under the pressure is pointed out to be caused by the enhancement of the FM coupling by the lattice contraction. Therefore, the suppression of $T_\text{N}$ by the Ca doping indicates that the dilution effect of the Eu moments by the Ca substitution exceeds the enhancement effect by the lattice contraction. It should be noted that the monotonic suppression of $T_\text{N}$ by the Ca doping is in contrast to the non-monotonic change of the electric transport (Fig. 2(a)). This different dependence might imply that the local electronic state, which is important to realize an axion insulator, remains unchanged except the dilution effect of the Eu moments, whereas the scattering effects on the conduction electrons are sensitively affected by an inhomogeneity of Ca doping.

With the magnetic effects of Ca substitution mapped, we next look for the Ca-doping effect on the topological physics via the magneto-transport properties. Figure 3 shows the field dependence of the magnetization $M$, the longitudinal magnetoresistance MR (defined as $MR = [\rho_{xx}(\mu_0 H) - \rho_{xx}(0)]/\rho_{xx}(0)$, and the Hall resistivity $\rho_{xy}$ in Ca$_x$Eu$_{1-x}$In2As2 with different $x$ at different temperatures under the magnetic field applied along the $c$ axis. As shown in the top panels in Fig. 3, the field dependence of $M$ shows a saturation above $\mu_0 H_\text{sat} \sim 2$ T with the saturation moment $M_\text{sat}$ that monotonically decreases with increasing the Ca ratio $x$, showing the dilution effect on the Eu moments by the Ca doping.

In contrast to the positive MR observed in the pristine sample below the saturation field region [11, 12, 20], the magnetic field dependence of the resistance of all the Ca-doped samples is dominated by the negative MR. This positive-to-negative MR observed in the pristine sample is discussed in terms of a crossover from weak localization to weak antilocalization due to the competition between the AFM and the FM interactions [25, 26], which might be affected by the Ca doping.

The evolution of $\rho_{xy}$ by the Ca doping shows a shrinking of the hole pocket of the Fermi surface of EuIn2As2. As shown in Figure 3(c), (f), (i) and (l), $\rho_{xy}$ of all the Ca-doped samples exhibit the positive slope, showing that the dominant carrier is still hole for all the samples. The nonlinear field dependence of $\rho_{xy}$ in the low-field region (Fig. 4) indicates an anomalous contribution in $\rho_{xy}$, which will be discussed later. From a linear fitting of the high-field data (see Fig. 4 for the data of $x = 0.06$ and $T = 2$ K), the hole carrier density ($n$) is estimated by the ordinary Hall component, $\rho_{xy}^0 = R_0 \mu_0 H$, where the ordinary Hall coefficient $R_0 = (ne)^{-1}$. As shown in Fig. 5, $n$ decreases with increasing Ca doping, indicating that the isovalent Ca-doping reduces the size of the hole pocket of EuIn2As2. This shrinking of the hole pocket can be simply understood by a gradual increase of the Fermi energy due to the lattice contraction in the Ca-doped samples.

To further confirm the lattice contraction effect on the band structure by the Ca doping, we performed first-principles calculations of the Ca-doped samples by creating a $2 \times 2$ supercell with a band folding (Fig. 6(a)). Here, the magnetic moments of the AFM phase are assumed to be oriented along the *a*-axis (Fig. 6(a)). We confirmed that the band structure obtained by our calculations does not sensitively depend on the arrangements of the Ca/Eu distribution except small energy differences (10–30 meV) by performing these calculations in the different arrangements of the Ca/Eu distribution (see Fig. S4 and Fig. S5 in the Supplementary Information). As shown in Fig. 6(b), the band structure we calculated for the pristine sample is in good agreement with the previous reports with the axion insulator phase [8, 10, 20]. We find that the bottom of the conduction bands around the *M* point is lowered to the Fermi level by introducing Ca atoms (Fig. 6(c, d)), indicating an increase of the Fermi level due to the contraction of the crystal structure by Ca doping. At the same time, we find that the calculated total density of states (DOS) at the Fermi level increases by increasing $x$ (Fig. 7), showing an electron doping by the lowering of the bands around the *M* point. Given the presence of the hole band in the actual EuIn$_2$As$_2$ owing to a lower Fermi level than that in the calculations, the increase of the Fermi level and the effective electron doping shown by our first-principles calculations are consistent with decrease of the hole carrier observed in our Hall measurements. We note that the shift of the peak in DOS to a higher energy is opposite to what would be expected from the increase of the Fermi level, which might be caused by a change of the band dispersion by Ca doping.

After analysis of the evolution of the Fermi level by estimating the change of carrier density, we turn to the anomalous part of the Hall resistivity. Generally, the Hall resistivity $\rho_{xy}$ can be expressed by:

$$\rho_{xy} = \rho_{xy}^O + \Delta\rho_{xy} = R_0\mu_0 H + \rho_{xy}^A + \rho_{xy}^T$$

By subtracting the ordinary Hall resistivity $\rho_{xy}^O$, the additional Hall resistivity $\Delta\rho_{xy}$ can be obtained as shown in Fig. 8 in different *x* at different temperatures. For an AHE in FM metals, the anomalous Hall resistivity $\rho_{xy}^A$ linearly scales with $M$ as $\rho_{xy}^A = R_s\mu_0 M$, where $R_s$ is the anomalous Hall coefficient. We determine $R_s$ ($R_s < 0$) by the field dependence of $\rho_{xy}^A$ above $\mu_0 H_{sat}$ and plot the field dependence of $R_s\mu_0 M$ as solid lines in Fig. 8. Firstly, $\rho_{xy}$ exhibits an anomalous Hall behavior even above $T_N$, which is same with pristine EuIn$_2$As$_2$ [11]. Same behaviors are observed in similar compounds EuZn$_2$As$_2$ [27] and both AFM- and FM-EuCd$_2$As$_2$ [28, 29], which can be attributed to Berry curvature anomalies in the momentum space. Furthermore, we fitted $\rho_{xy}^A$ data to the $\rho_{xx}$ value of Ca$_x$Eu$_{1-x}$In$_2$As$_2$ to understand the origin of the anomalous Hall effect (intrinsic or extrinsic) in the present system. As shown in Fig. S6, we find a linear relation between $\rho_{xy}^A$ and $\rho_{xx}^2$ in the low-temperature region, supporting the intrinsic Berry curvature mechanism for the AHE in this system.

Apart from $\rho_{xy}^A$, a clear deviation from $\Delta\rho_{xy} \propto M$ is observed in $\Delta\rho_{xy}$ for $H < H_{sat}$ and $T < T_N$ as shown as the shaded area in Fig. 8. Given the linear scaling of $\rho_{xy}^A \propto M$ for either intrinsic or extrinsic anomalous Hall effects, this deviation demonstrates an additional component beside the anomalous Hall effect in $\Delta\rho_{xy}$, named as a topological (or geometrical) Hall effect (THE) $\rho_{xy}^T$ as done in the previous work [11]. This THE is only observed below $T_N$ for all Ca ratio, indicating that the THE is tightly related to the AFM ordering in Ca$_x$Eu$_{1-x}$In$_2$As$_2$. It should be noted that all the Ca-doped samples show a similar field dependence of the THE despite the non-monotonic change of $\rho_{xx}$ in the different Ca doping (Fig. 2(a)). This different dependence of the THE and $\rho_{xx}$ on the Ca ratio shows that, whereas the scatterings of the conduction electrons might be sensitively affected by inhomogeneity of the Ca concentration, the magnetic state of Eu moments is stable against the inhomogeneity. In contrast to the THE observed even above $T_N$ in the similar A-type AFM EuZn$_2$As$_2$ [27], the THE in the Ca-doped samples disappears above $T_N$, but AHE persists above $T_N$, indicating that the origin of THE is distinct in the two systems due to the different magnetic structures [20]. In the pristine EuIn$_2$As$_2$, the THE is pointed to originate from the field-induced spin chirality in the real space due to the helical magnetic ground state [11, 20]. Therefore, the persistence of the THE in our Ca-doped EuIn$_2$As$_2$ samples shows that the magnetic structure of the pristine EuIn$_2$As$_2$ might remain unchanged by the Ca doping.

In summary, we systematically investigated the magnetic and the transport properties in a series of Ca$_x$Eu$_{1-x}$In$_2$As$_2$ via both the experimental measurements and the first-principles calculations. We find the decrease of both the *a*- and *c*-axis lengths as increasing the Ca doping ratios. Our Hall measurements reveal that the isovalent Ca doping effectively reduces the hole band by the lattice contraction, which is confirmed by first-principles calculations. Most importantly, we find that both the temperature dependence of the magnetic susceptibility with the local maximum at the Néel temperature and the field dependence of the topological Hall effect are not changed in the Ca-doped EuIn$_2$As$_2$, except the gradual decrease of the Néel temperature by the dilution effect of the Eu moments. These results indicate that we can tune the Fermi level of EuIn$_2$As with keeping the magnetic structure by the Ca doping, demonstrating that the Ca-doped EuIn$_2$As$_2$ is an ideal system to realize the axion insulating state by controlling the carrier density without changing the magnetic structure.

**Acknowledgements**
We thank Dr. Daisuke Hamane for performing the EDS experiments using facilities of the Institute for Solid State Physics, the University of Tokyo and helpful discussion. J.Y. was supported by Grant-in-Aid for JSPS Fellows. J.Y. and M.Y. thank the support by JSPS KAKENHI Grant Numbers JP22KF0111 and JP23H01116. J. G. S and B.-T. W acknowledge funding from National Natural Science Foundation of China (Grant No. 12304095) and the Guangdong Basic and Applied Basic Research Foundation (Grant No. 2023A1515140188). X. L. and Y. P. S. thank the support from the National Key R&D Program (Grant Nos. 2023YFA1607402, 2021YFA1600201), the National Natural Science Foundation of China (Grant Nos. U2032215, No. U1932217, and 12274412) and Systematic Fundamental Research Program Leveraging Major Scientific and

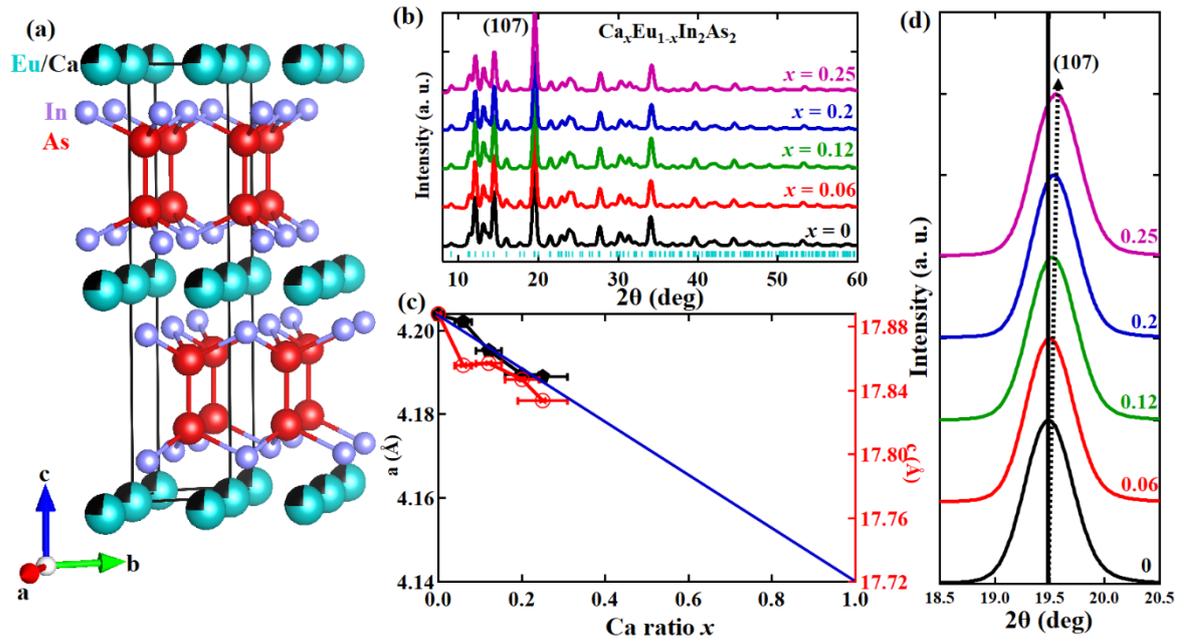

Figure 1. (a) Schematic crystal structure of Ca-doped EuIn$_2$As$_2$. (b) X-ray diffraction (XRD) patterns of Ca$_x$Eu$_{1-x}$In$_2$As$_2$ with $0 < x < 0.37$. The cyan ticks denote Bragg peak positions of EuIn$_2$As$_2$ with space group $P6_3/mmc$ (No. 196) (see Fig. S2 in Supplementary Information for more details). (c) Evolution of the lattice parameters $a$ and $c$ with varying Ca ratio $x$ with vertical (smaller than the symbol size) and horizontal (determined by the deviation in the EDS data) error bars. The blue line indicates the extrapolation of lattice parameter from EuIn$_2$As$_2$ to CaIn$_2$As$_2$. (d) Zoom-in view of the most intense diffraction peak (107) for different $x$. The dashed line indicates the peak shift to the high-angle direction with increasing $x$ from the peak position at $x = 0$ (the vertical dotted line).

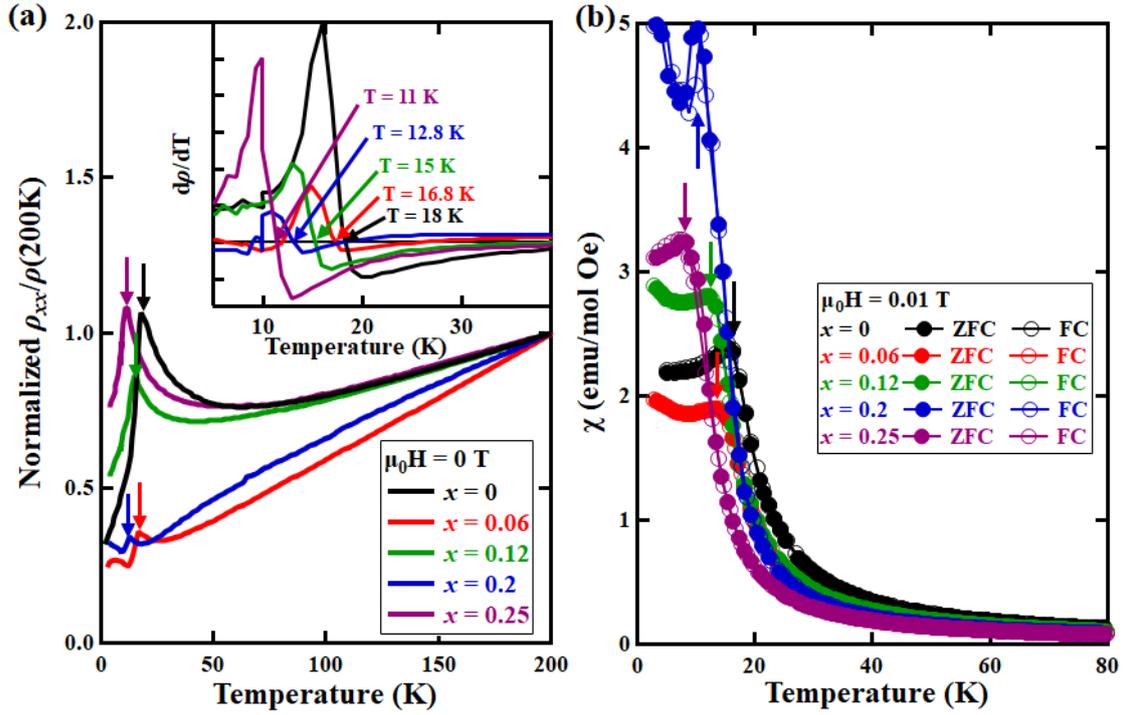

Figure 2. (a) Normalized electrical resistivity $\rho_{xx}$ and (b) Temperature dependence of magnetic susceptibility $\chi$ for $Ca_xEu_{1-x}In_2As_2$ under ZFC and FC modes with applied magnetic field $\mu_0H = 0.01$ T. The arrows indicate the AFM transition temperatures for $Ca_xEu_{1-x}In_2As_2$. The inset of (a) shows the temperature dependence of the derivative of the electrical resistivity $d\rho/dT$.

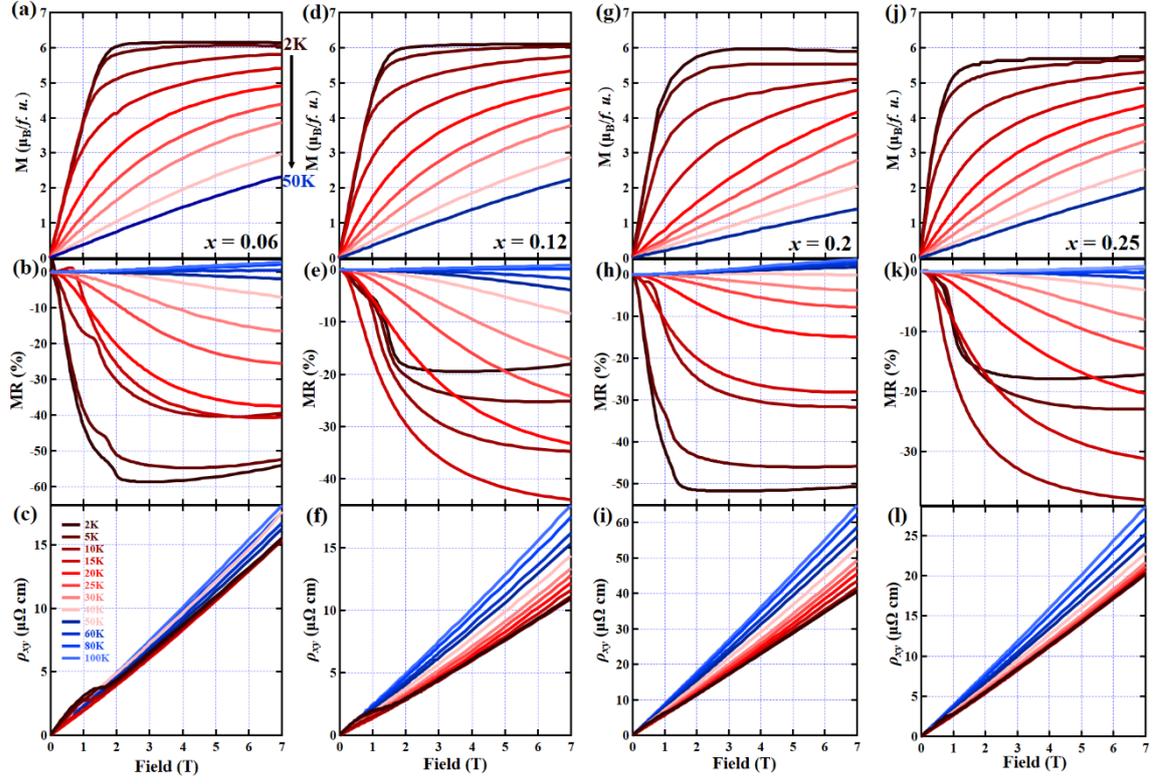

Figure 3. The magnetization (*M*), the magnetoresistance (MR) and the Hall resistivity $\rho_{xy}$ as a function of magnetic field applied along the *c* axis for (a-c) $x$ = 0.06, (d-f) $x$ = 0.12, (g-i) $x$ = 0.2, (j-l) $x$ = 0.25 in $Ca_xEu_{1-x}In_2As_2$. The different colors of each line indicate the measured temperatures of each data.

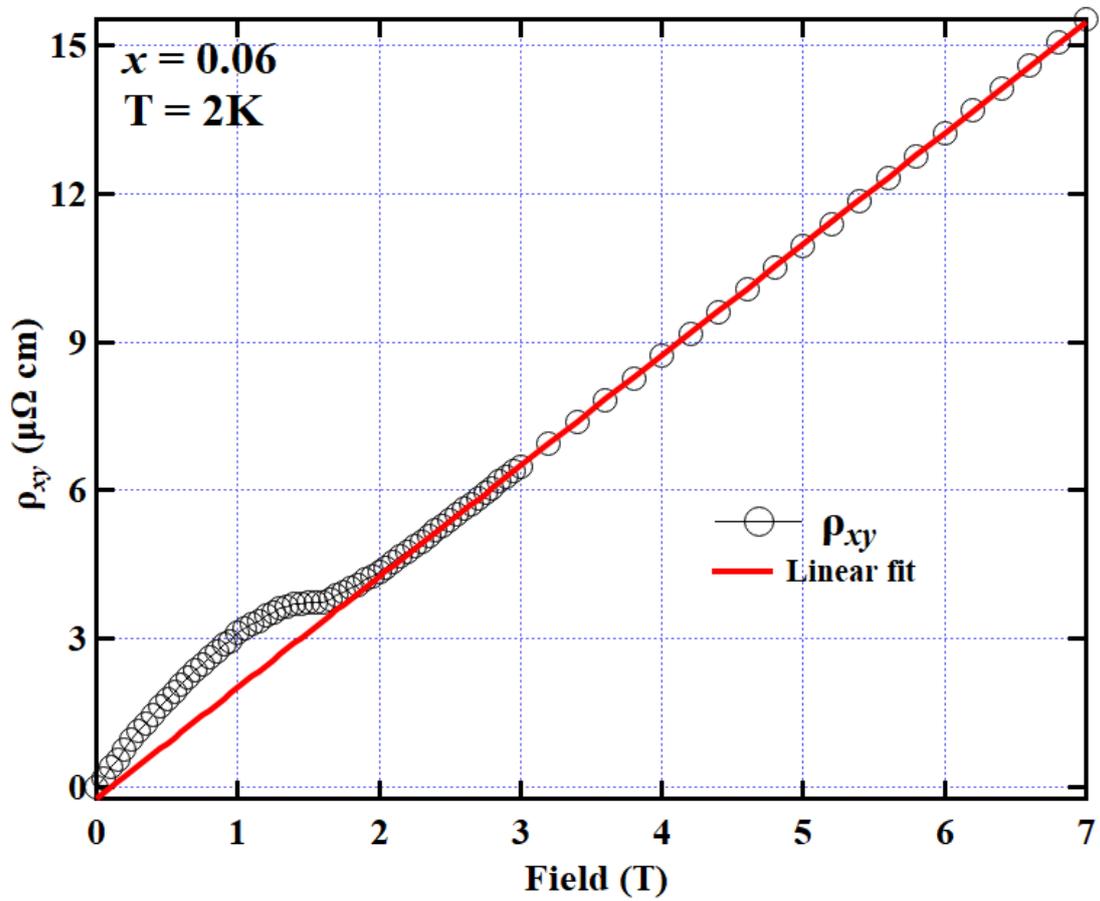

Figure 4. The magnetic field dependence of the Hall resistivity $\rho_{xy}$ at 2 K for $x = 0.06$. The red solid line shows a linear fit to the data above the saturation field, which is used to estimate the ordinary Hall coefficient $R_0$ and the anomalous Hall coefficient $R_s$ from the slope and the intercept of the fit, respectively.

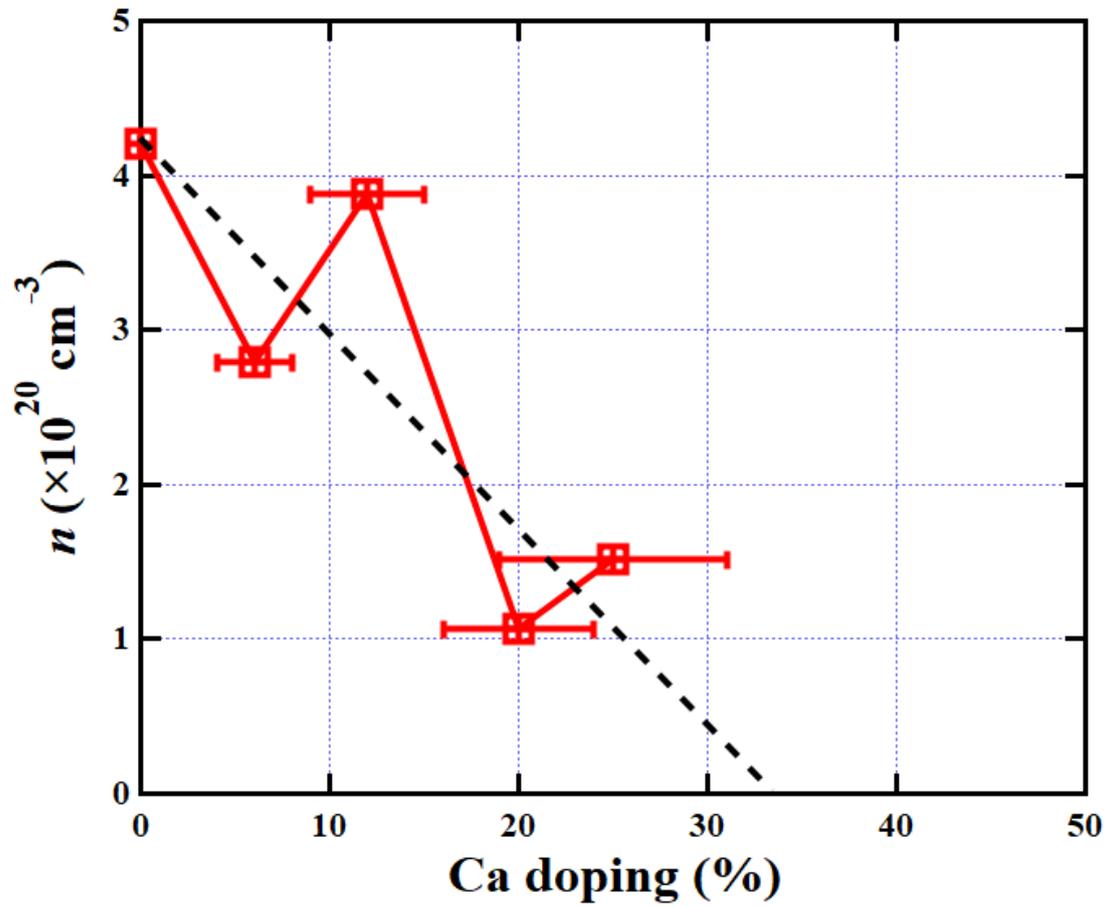

Figure 5. The carrier density *n* as a function of Ca doping ratio. The black dotted line is the guide for the eye.

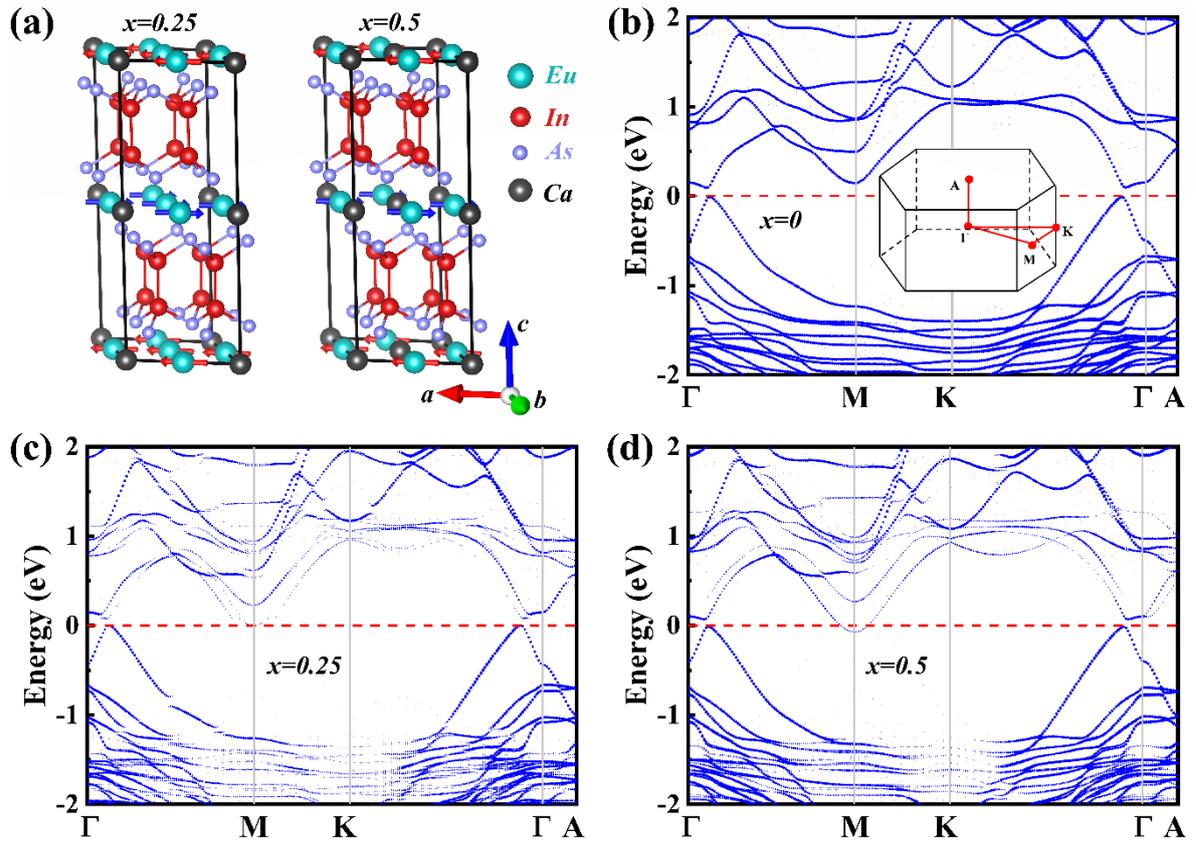

Figure 6. (a) schematic illustrations of the crystal and the magnetic structure of Ca-doped EuIn$_2$As$_2$. The blue and red arrows at the Eu sites show the magnetic structure used in the calculations. (b-d) Electronic band structures along the high-symmetry path in Brillouin zone (BZ) of the pristine $x = 0$ (b), $x = 0.25$ (c), and $x = 0.5$ doped (d) EuIn$_2$As$_2$ with the A-type antiferromagnetism and spin-orbit coupling. The inset in (b) shows the bulk BZ exhibiting the highly symmetric paths used to calculate the energy bands.

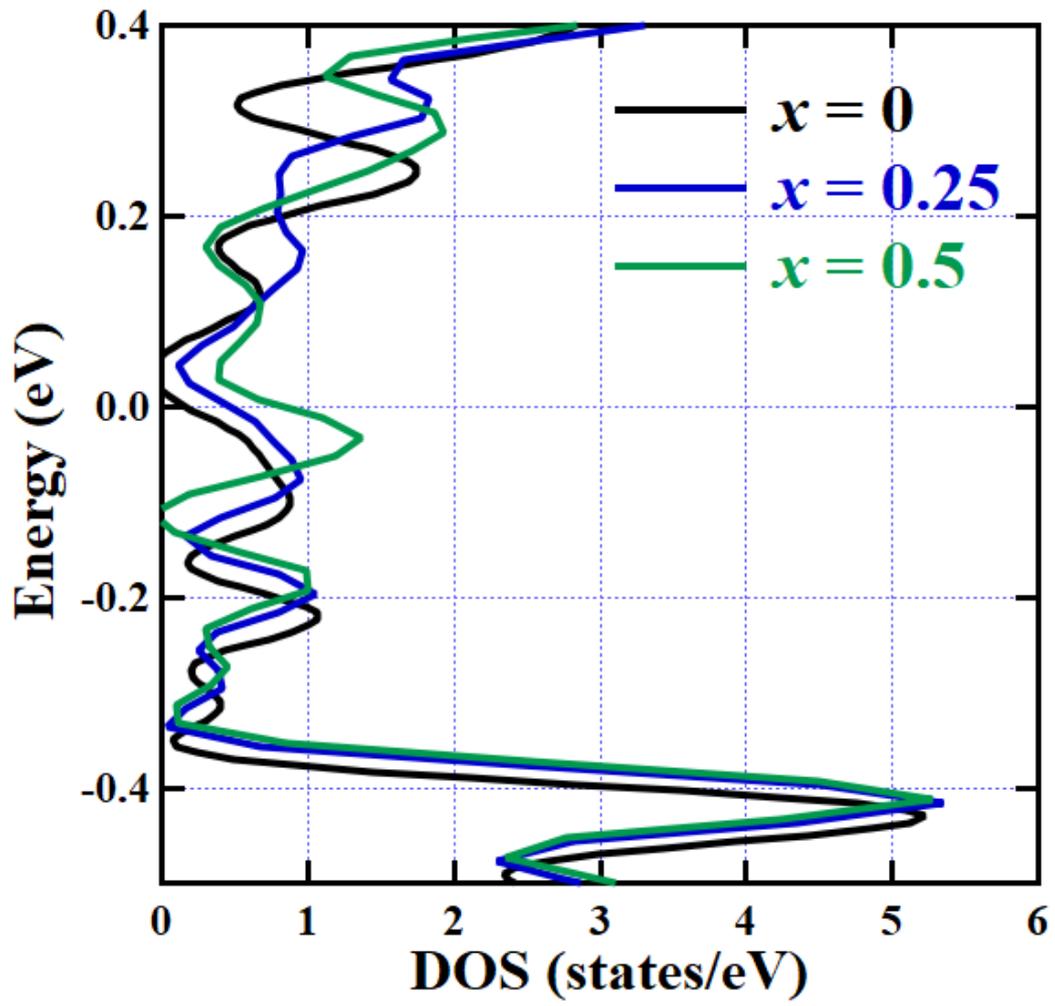

Figure 7. The total DOS of the different chosen Ca ratios in our calculation.

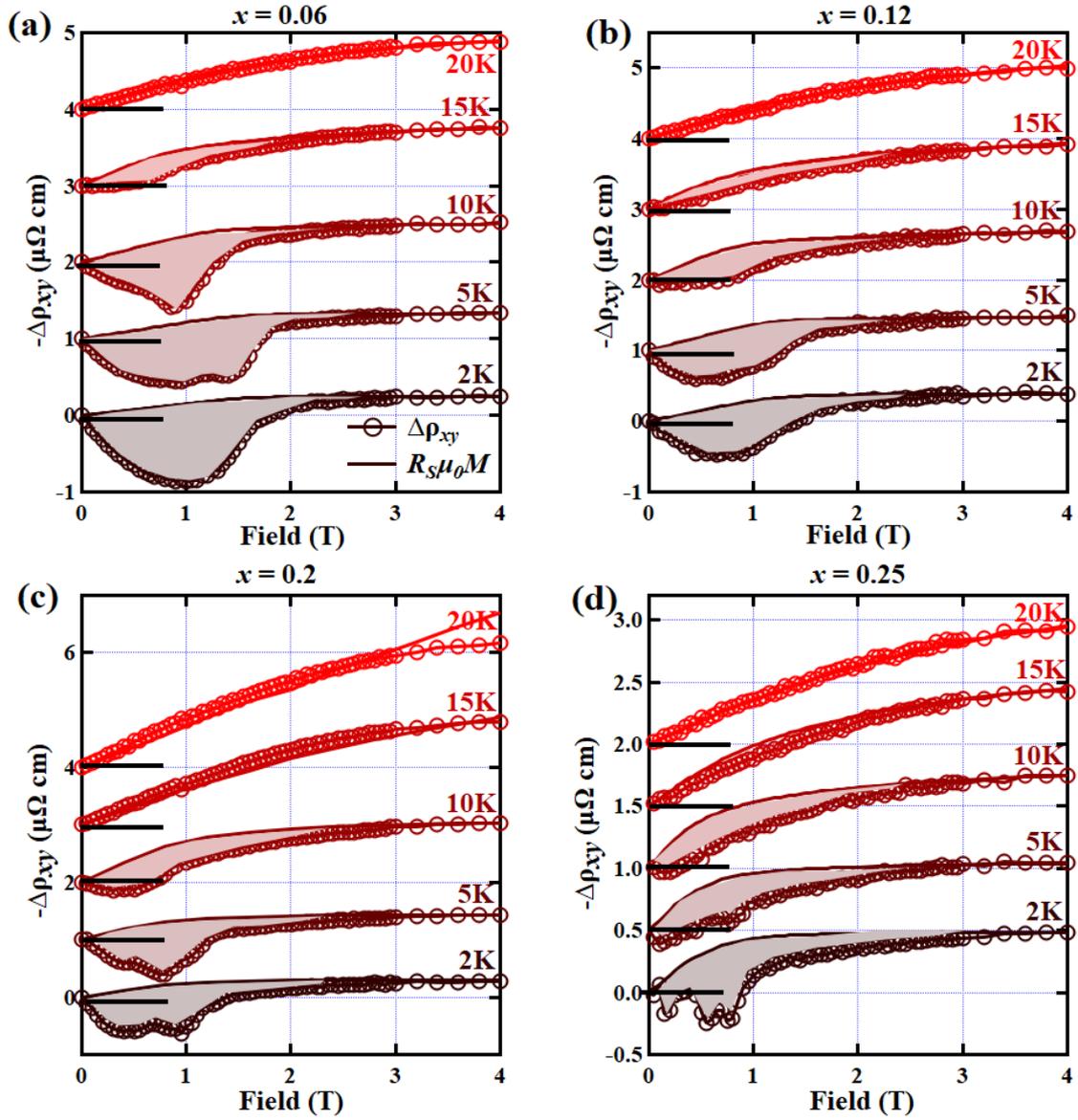

Figure 8. The magnetic field dependence of the additional Hall resistivity $\Delta\rho_{xy}$ (circles) and $R_S\mu_0 M$ (solid lines) at different temperatures for different Ca ratio $x$, (a) $x = 0.06$; (b) $x = 0.12$; (c) $x = 0.2$; (d) $x = 0.25$. The data are vertically shifted for clarity. The topological Hall components are marked by the gray shaded regions.